\documentclass[12pt,preprint]{aastex}

\slugcomment{}

\shorttitle{3c138}
\shortauthors{Shen et al.}

\begin{document}

%% LaTeX will automatically break titles if they run longer than
%% one line. However, you may use \\ to force a line break if
%% you desire.

\title{The Center of Activity in the CSS Superluminal Source 3C 138}

%% Use \author, \affil, and the \and command to format
%% author and affiliation information.
%% Note that \email has replaced the old \authoremail command
%% from AASTeX v4.0. You can use \email to mark an email address
%% anywhere in the paper, not just in the front matter.
%% As in the title, use \\ to force line breaks.

\author{Z.-Q. Shen\altaffilmark{1}, L.-L. Shang\altaffilmark{1,2}, H.-B. Cai\altaffilmark{1,2},
X. Chen\altaffilmark{1,2}, D. R. Jiang\altaffilmark{1}, Y.-J. Chen\altaffilmark{1}, X.
Liu\altaffilmark{3}, R. Yang\altaffilmark{4}, S.
Kameno\altaffilmark{5} and H. Hirabayashi\altaffilmark{6}}
%\affil{Shanghai Astronomical Observatory, Chinese Academy of
%Sciences, Shanghai 200030, P. R. China}

%\author{C. D. Biemesderfer\altaffilmark{4,5}}
%\affil{National Optical Astronomy Observatories, Tucson, AZ 85719}
%\email{aastex-help@aas.org}

%\and

%\author{R. J. Hanisch\altaffilmark{5}}
%\affil{Space Telescope Science Institute, Baltimore, MD 21218}

%% Notice that each of these authors has alternate affiliations, which
%% are identified by the \altaffilmark after each name.  Specify alternate
%% affiliation information with \altaffiltext, with one command per each
%% affiliation.

\altaffiltext{1}{Shanghai Astronomical Observatory, Chinese
Academy of Sciences, Shanghai 200030, P. R. China}
\altaffiltext{2}{Graduate School of Chinese Academy of Sciences,
Beijing 100039, P. R. China} \altaffiltext{3}{Urumqi Observatory,
National Astronomical Observatories, Urumqi 830011, P. R. China}
\altaffiltext{4}{Department of Physics, Fudan University, Shanghai
200433, P. R. China} \altaffiltext{5}{National Astronomical
Observatory, Mitaka, Tokyo 181-8588, Japan}
\altaffiltext{5}{Institute of Space and Astronautical Science
(ISAS)/JAXA, Sagamihara, Kanagawa 229-8510, Japan}

%% Mark off your abstract in the ``abstract'' environment. In the manuscript
%% style, abstract will output a Received/Accepted line after the
%% title and affiliation information. No date will appear since the author
%% does not have this information. The dates will be filled in by the
%% editorial office after submission.

\begin{abstract}
We present the results from the first quasi-simultaneous
multi-frequency (2.3, 5.0, 8.4 and 15 GHz) Very Long Baseline
Interferometry (VLBI) observations of a compact steep spectrum
(CSS) superluminal source 3C138.  For the first time, the spectral
distribution of the components within its central 10
milli-arcsecond (mas) region was obtained. This enables us to
identify the component at the western end as the location of the
nuclear activity on the assumption that the central engine is
associated with one of the detected components. The possibility
that none of these visible components is the true core is also
discussed. The new measurements further clarify the superluminal
motions of its inner jet components. The multi-frequency data
reveal a convex spectrum in one jet component, inferring the
existence of free-free absorption by the ambient dense plasma.
\end{abstract}

%% Keywords should appear after the \end{abstract} command. The uncommented
%% example has been keyed in ApJ style. See the instructions to authors
%% for the journal to which you are submitting your paper to determine
%% what keyword punctuation is appropriate.

%% Authors who wish to have the most important objects in their paper
%% linked in the electronic edition to a data center may do so in the
%% subject header.  Objects should be in the appropriate "individual"
%% headers (e.g. quasars: individual, stars: individual, etc.) with the
%% additional provision that the total number of headers, including each
%% individual object, not exceed six.  The \objectname{} macro, and its
%% alias \object{}, is used to mark each object.  The macro takes the object
%% name as its primary argument.  This name will appear in the paper
%% and serve as the link's anchor in the electronic edition if the name
%% is recognized by the data centers.  The macro also takes an optional
%% argument in parentheses in cases where the data center identification
%% differs from what is to be printed in the paper.


%% From the front matter, we move on to the body of the paper.
%% In the first two sections, notice the use of the natbib \citep
%% and \citet commands to identify citations.  The citations are
%% tied to the reference list via symbolic KEYs. The KEY corresponds
%% to the KEY in the \bibitem in the reference list below. We have
%% chosen the first three characters of the first author's name plus
%% the last two numeral of the year of publication as our KEY for
%% each reference.

\section{Introduction}

The compact steep spectrum (CSS) source 3C138 ($m_{v}$ = 18.84;
z=0.759) is a powerful quasar. It has a convex spectrum peaked at
$\sim$130 MHz with a steep high frequency spectrum of 0.65
($S_{\nu}\propto\nu^{\alpha}$), suggesting that its total flux
density is dominated by the emissions from the jet and lobe
components. Its arc second-scale radio structure consists of a
core, several bright jet knots, a compact lobe to the east and a
fainter, more diffuse lobe to the west \citep[and references
therein]{aku93}. High resolution VLBI observations
\citep{she01,cot03} have revealed at least three compact
components within the central core region. The exact location of
the real core in 3C 138, however, remains uncertain. This is
mainly due to the lack of high-resolution spectral information on
two possible candidates \citep[components A and B in][]{cot03}.
From their earlier brightness measurements of two components,
\citet{fan89} claimed that component B, which is bright and
compact, would be associated with the central engine. Based on the
very weak linear polarization ($<$0.4\%) in component A compared
to a peak polarized intensity of 3.5\% for component B at 5.0 GHz
\citep{cot97} and the less time variability in component B over
$\approx$12 years \citep{she01}, it is argued that the western end
component A is most likely the location of the nuclear activity.

In this paper, we will present the results on the core
identification using the spectral data obtained from the first
quasi-simultaneous multi-frequency VLBI imaging of 3C 138.

\section{Observations and data reduction}

%% In a manner similar to \objectname authors can provide links to dataset
%% hosted at participating data centers via the \dataset{} command.  The
%% second curly bracket argument is printed in the text while the first
%% parentheses argument serves as the valid data set identifier.  Large
%% lists of data set are best provided in a table (see Table 3 for an example).
%% Valid data set identifiers should be obtained from the data center that
%% is currently hosting the data.

The observations were performed with the NRAO\footnote{The
National Astronomy Observatory (NRAO) is operated by Associated
Universities Inc., under cooperative agreement with the National
Science Foundation.} Very Long Baseline Array (VLBA) on August 20,
2001. With the capability of the frequency switching, 3C 138 was
observed at four frequency bands including the dual frequency
(2.3/8.4 GHz), 5.0 GHz and 15.4 GHz. Observations at different
frequencies were interlaced to ensure the comparable $(u,v)$
coverage at each band, with a total observing time of 130, 130 and
250 minutes at dual 2.3/8.4, 5.0 and 15.4 GHz, respectively. For
each scan, data were recorded in 1 bit sampling VLBA format with a
total bandwidth of 64 MHz (eight 8 MHz IF channels) per circular
polarization at each station. For the dual frequency (2.3/8.4 GHz)
scans, the right-circular polarized (RCP) radio signals were
recorded simultaneously with four 8 MHz channels for each of 2.3
and 8.4 GHz. The left-circular polarized (LCP) signals were
recorded in all 8 IFs for scans at both 5.0 and 15.4 GHz.

The data correlation was made at the VLBA correlator in Socorro,
New Mexico, USA. All of the post-correlation data reduction was
carried out within the NRAO AIPS \citep{sch83} software and the
Caltech DIFMAP \citep{she97} package. A priori visibility
amplitude calibration was done using the antenna gain and the
system temperature measured at each station. The global fringe
fitting was successfully performed for observations at four
frequencies of 2.3, 5.0, 8.4 and 15.4 GHz. To minimize the
smearing effects on the large field ($\approx$400 mas) imaging,
fringe-fitted data were averaged to 20 sec in time and kept 8 MHz
of each IF in frequency in the process of the commonly used
self-calibration iteration. As a result, high-resolution VLBI
images of 3C 138 (including both the extended emission from the
hotspots and jet knots and the compact emission from the central
core region) were made at frequencies 2.3, 5.0 and 8.4 GHz (Fig.
1). The first 15.4 GHz VLBI image of 3C 138 only shows the central
compact core emission (Fig. 2) because the large scale structure
is heavily resolved.

\section{Results}

\subsection{structure}

It can be seen from Fig. 1 (left) that the large scale structure
of 3C 138 observed at three different frequencies is very similar.
It consists of two distinct emission regions at two ends which are
separated by $\approx$400 mas along a position angle of 70$^{\rm
o}$ and, some discrete jet knots seen in between. These jet knot
components are heavily resolved with more diffuse emissions
recovered at lower frequency. These agree very well with the
existing VLBI images. The misplacement and missing of some knots
at one or two frequencies are treated as artifacts mainly due to
the complex structure of the source emission at this scale. The
counter-jet emission at $\approx$250 mas west to the compact core
seen by 1.7 GHz VLBI observations \citep{cot97} was not detected
in our observations, consistent with the non-detection results
from past 5.0 GHz VLBI observations \citep{she01,cot03}.

In the central 10 mas core region, in addition to the previously
reported three components \citep[e.g.][]{she01}, a new component
was consistently seen at 5.0, 8.4 and 15.4 GHz in August 2001
(Figs. 1 and 2). This can be identified with the component B2
appearing on the 5.0 GHz linear polarization images \citep{cot03}
in three epochs from September 1998 to October 2002.

\subsection{Spectrum and core identification}

For the first time, four central components A, B1, B2 and C
\citep[after][]{cot03} were seen at four frequencies
quasi-simultaneously. This removes any time variation in the
structure and makes it possible to estimate the component's
spectral index which can be used to clarify the core
identification. The quantitative description of the source
structure in the central 10 mas was determined by model fitting to
the calibrated visibility data at each frequency. The results are
tabulated in Table 1. The first column is the component
designation followed by the component's flux density in Jy,
separation and position angle of each component with respect to
component A in mas and degrees, respectively, the size (FWHM) of
circular Gaussian component in mas, and the component's brightness
temperature in the source rest frame estimated using the formula
given in \citet{shen97}.

Fig. 3 shows the spectral distribution based on these
measurements. The fitting results of spectral index $\alpha$
(S$_\nu$ $\propto$ $\nu$$^{-\alpha}$) are listed in Table 2. The
uncertainties of the spectral indices are the errors of fitting to
the modelled flux densities in Table 1. Since components B1 and B2
were not well resolved at 2.3 GHz (see Fig. 1) and there is a
significant absorption towards component C at 2.3 GHz (see below),
only measurements at 5.0, 8.4 and 15.4 GHz are used in the least
squares fitting for the spectral indices ($\alpha_{\rm 5.0 \phd
GHz}^{\rm 15.4 \phd GHz}$) of components B1, B2 and C. For
components A and B (as the combination of components B1 and B2 at
frequencies higher than 2.3 GHz), spectral indices between 2.3 and
15.4 GHz ($\alpha_{\rm 2.3 \phd GHz}^{\rm 15.4 \phd GHz}$) are
also estimated as well as $\alpha_{\rm 5.0 \phd GHz}^{\rm 15.4
\phd GHz}$. It can be seen that component A has the flattest
spectral index $\alpha_{\rm 5.0 \phd GHz}^{\rm 15.4 \phd GHz}$ of
0.30. Adding one more data point at 2.3 GHz in the fitting, we
obtained the fitted spectral index $\alpha_{\rm 2.3 \phd GHz}^{\rm
15.4 \phd GHz}$ of 0.37 for component A. Again, this is flatter
than that of 0.59 for another candidate (component B). This is in
favor of component A being the nuclear component. Both components
B1 and B2 have almost the same spectral indices of 0.57 and 0.54,
suggesting the same origin of or environment in both components.
Thus, it is very unlikely that either of B1 and B2 was the
location of the central engine. Our core identification is further
supported by the measured brightness temperature T$_{\rm B}$ (see
Table 1). Component A has the highest T$_{\rm B}$ of
3.4$\times$10$^9$ K at 15.4 GHz which is almost kept unchanged at
other lower frequencies, while both components B1 and B2 see a
decreasing in T$_{\rm B}$ with the frequency. This identification
is consistent with the results from the studies of the linear
polarization \citep{cot97,cot03} and variability \citep{she01}.
Component C is a typical jet component with the spectral index of
1.15 between 5.0 and 15.4 GHz. For comparison, we also calculated
the spectral index $\alpha_{\rm 2.3 \phd GHz}^{\rm 8.4 \phd GHz}$
for both lobe emission at the eastern end of $\approx$400 mas and
a jet emission at $\approx$25 mas east to the central core region
(see Fig. 1) based on their integrated flux densities estimated
from the images at 2.3, 5.0 and 8.4 GHz . They are eventually the
same of about 1.5.

To ensure an accurate absolute flux density calibration at all the
observing frequencies, a strong compact quasar PKS 0528+134 was
observed as a flux density calibrator during our VLBA observations
of 3C 138. The comparison between the total flux density
measurements at 5.0, 8.4 and 15.4 GHz by the University of
Michigan Radio Astronomy Observatory (UMRAO) and at 2.3 GHz by the
NRAO Green Bank Interferometer (GBI), and the integrated flux
densities in the VLBA images, indicates that the errors in the
absolute flux density calibration are about 3\%, 5\%, 2\% and 10\%
at frequency of 2.3, 5.0, 8.4 and 15.4 GHz, respectively (Cai et
al. in preparation). In addition, there are typical 10\% errors in
the flux density due to the Gaussian model fitting. Taking into
account the overall error budgets due to these effects in the
spectral fitting, we obtained $\alpha_{\rm 2.3 \phd GHz}^{\rm 15.4
\phd GHz}$ for components A and B of 0.38$\pm$0.09 and
0.60$\pm$0.09, respectively, supporting the core identification
discussed above.

For component C, there must be an absorption at 2.3 GHz in order
to fit in the spectral shape between 4.8 and 15.4 GHz. The
mechanism for the absorption could be the intrinsic Synchrotron
Self-Absorption (SSA) or the Free-Free Absorption (FFA) by the
ambient cold plasma. Both can produce convex spectra: S$_\nu$
$\propto$ $\nu$$^{2.5}$[1 -- exp(-$\tau_{\rm
s}$$\nu^{-(2.5+\alpha)})]$ for SSA, and S$_\nu$ $\propto$
$\nu$$^{-\alpha}$ exp(-$\tau_{\rm f}$$\nu^{-2.1})$ for FFA, where
$\nu$ is the observing frequency in GHz, $\tau_{\rm s}$ and
$\tau_{\rm f}$ are the SSA and FFA coefficients at 1 GHz,
respectively, and $\alpha$ is the spectral index. Either of two
models (SSA and FFA) can fit the observed convex spectrum of
component C quite well. With a fixed spectral index $\alpha$=1.2
(see Table 2), the fitted 1 GHz absorption coefficients and
Synchrotron flux density are $\tau_{\rm s}$=32 and S$_0$=8.6 mJy
and, $\tau_{\rm f}$=4.3 and $S_0$=300 mJy for SSA and FFA,
respectively. For SSA model, the fitted spectrum has a peak flux
density of 59 mJy at a turn-over frequency 2.8 GHz. This would
require a very large magnetic field within component C of about 35
G, which is unrealistic for component C to maintain its
synchrotron emission for years against synchrotron loss
\citep{kel81}. However, the non-detection of any polarized
emission in the inner jet component C is consistent with component
C being surrounded by a patchy but dense medium \citep{cot03}.
Thus, it is very likely that the observed absorption in component
C is mainly due to FFA by the ambient cold dense plasma.

\subsection{Proper motion}

Previous studies have detected superluminal motions in central
core region of 3C 138 \citep{she01,cot03}. To avoid any possible
position offset among the measurements made at different
frequencies, only 5.0 GHz data (including our new observations)
were used to refine the proper motion calculations. Data points at
epochs 1985.50, 1989.72 and 1997.85 are from \citet[]{she01}. Data
points at epochs 1994.97, 1997.60, 1998.70, 2000.59 and 2002.79
are from \citet[]{cot03}. The data point at epoch 2001.64 is from
this work which has well resolved component B into components B1
and B2. By assuming that the position of component B is at the
weighted  center of two components (B1 and B2) system with the
weight proportional to their flux densities, we can obtain the
position of component B from the model fitting results of
components B1 and B2 at epoch 2001.64.

As shown in Fig. 4, there is no significant change in the speed of
motion in both components B (in $\approx$18 years) and C (in
$\approx$8 years) with regard to component A which is assumed to
be stationary as the location of the nucleus. The best fitted
proper motion is 0.072$\pm$0.010 and 0.20$\pm$0.03 mas yr$^{-1}$,
corresponding to an apparent superluminal speed of 2.6$\pm$0.4 c
and 7.2$\pm$1.0 c for components B and C, respectively (assuming
H$_0$=65 km s$^{-1}$ Mpc$^{-1}$ and q$_0$=0.5). This is consistent
with the published estimate of 3.3 c for component B
\citep{she01}. Component C, which is farther away than component B
from the nucleus component A, has a much faster motion than
component B. This could be due to the (re-)acceleration or a
smaller viewing angle of its emission to the observer's line of
sight.

The new component B2 (see Figs. 1 and 2) was first clearly
detected in the linear polarization image in September 1998
\citep{cot03}. Since then, both components B1 and B2 have been
consistently seen in another two epochs of high-resolution
polarization sensitive VLBA observations by \citet{cot03}. Inset
of Fig. 4 is a plot of the separations between two components B1
and B2 as a function of the observing epochs. Three data points
denoted by open circles are estimated from the polarized images
\citep{cot03}, and the data point represented by a star is from
our total intensity measurement (Table 1).

Obviously, there is a systematic offset of the separation in
linear polarized emission from that of total intensity. This is
consistent with the significant offsets of the peaks in the
polarized and total intensities observed simultaneously, and can
be explained by the ``holes in a (Faraday) screen'' scenario
\citep{cot03}. Because of this, the total intensity measurement
was not used in the linear fit which gives a fitted proper motion
of --0.009$\pm$0.015 mas yr$^{-1}$ (represented by a dashed line).
Currently, this result is consistent with the claim that there is
no relative motion between components B1 and B2 \citep{cot03}
before any future new observations are available to decrease the
uncertainty in the determination of the relative motion. If
confirmed, such a proper motion would indicate that component B2
is moving towards component B1 at a relative speed of
0.32$\pm$0.54 c, which would be independent evidence that neither
B1 nor B2 is related to the location of central activity.

\section{Discussion}

The main results of this work are summarized in the abstract, so
we avoid the redundancy here. It should be noted that all the
arguments on the core identification are based on an underlying
assumption that central engine must be associated with one of the
detected components. But this may not be the case for CSS sources
whose emission is usually dominated by the strong knot/lobe/jet
emissions. It is proposed \citep{fan95,rea96} that CSS sources are
part of evolutionary sequence, in which they represents an early
stage between the compact symmetric objects and large
Fanaroff-Riley type II objects. The true core of 3C 138 which is a
prototype of CSS source, could be too weak to be seen, or simply
embedded in the surrounding dense medium. If so, the exact
location of the center of activity in 3C 138 still remains
undetected.

The phase-referenced VLBI observations can provide precise
positional information with respect to the external reference
source with a sub-milliarcsecond accuracy. By examining the
component's absolute proper motion relative to the more distant
compact reference source \citep{bar86}, one can confirm or exclude
the presumed location of the center of activity dynamically.

During our VLBA observations at 15.4 GHz, the phase referencing
technique was also adopted by fast switching between 3C 138 and a
nearby (in angular separation) but distant (z=2.07) bright compact
quasar PKS 0528+134, with an observing cycle time of 100 s,
consisting of 32 s on PKS 0528+134, 8 s for antenna slewing, 52 s
on 3C138, and another 8 s for antenna slewing. This was turned out
to be successful with components A and B are well detected
\citep{sha04}. With more epochs of observations in the future,
this will enable us to explore the possibility that none of these
visible components is the true core.

%\section{Conclusions}

%% The \notetoeditor{TEXT} command allows the author to communicate
%% information to the copy editor.  This information will appear as a
%% footnote on the printed copy for the manuscript style file.  Nothing will
%% appear on the printed copy if the preprint or
%% preprint2 style files are used.

%% The eqnarray environment produces multi-line display math. The end of
%% each line is marked with a \\. Lines will be numbered unless the \\
%% is preceded by a \nonumber command.
%% Alignment points are marked by ampersands (&). There should be two
%% ampersands (&) per line.

%% This section contains more display math examples, including unnumbered
%% equations (displaymath environment). The last paragraph includes some
%% examples of in-line math featuring a couple of the AASTeX symbol macros.

%% The displaymath environment will produce the same sort of equation as
%% the equation environment, except that the equation will not be numbered
%% by LaTeX.

%% If you wish to include an acknowledgments section in your paper,
%% separate it off from the body of the text using the \acknowledgments
%% command.

%% Included in this acknowledgments section are examples of the
%% AASTeX hypertext markup commands. Use \url without the optional [HREF]
%% argument when you want to print the url directly in the text. Otherwise,
%% use either \url or \anchor, with the HREF as the first argument and the
%% text to be printed in the second.

\acknowledgments

This research has made use of data from the UMRAO, which is
supported by funds from the University of Michigan. The Green Bank
Interferometer is a facility of the NSF operated by the NRAO in
support of NASA High Energy Astrophysics programs.

\clearpage

%% Use the figure environment and \plotone or \plottwo to include
%% figures and captions in your electronic submission.
%% To embed the sample graphics in
%% the file, uncomment the \plotone, \plottwo, and
%% \includegraphics commands
%%
%% If you need a layout that cannot be achieved with \plotone or
%% \plottwo, you can invoke the graphicx package directly with the
%% \includegraphics command or use \plotfiddle. For more information,
%% please see the tutorial on "Using Electronic Art with AASTeX" in the
%% documentation section at the AASTeX Web site,
%% http://www.journals.uchicago.edu/AAS/AASTeX.
%%
%% The examples below also include sample markup for submission of
%% supplemental electronic materials. As always, be sure to check
%% the instructions to authors for the journal you are submitting to
%% for specific submissions guidelines as they vary from
%% journal to journal.

%% This example uses \plotone to include an EPS file scaled to
%% 80% of its natural size with \epsscale. Its caption
%% has been written to indicate that additional figure parts will be
%% available in the electronic journal.

\begin{figure}
\includegraphics[angle=-90,scale=.70]{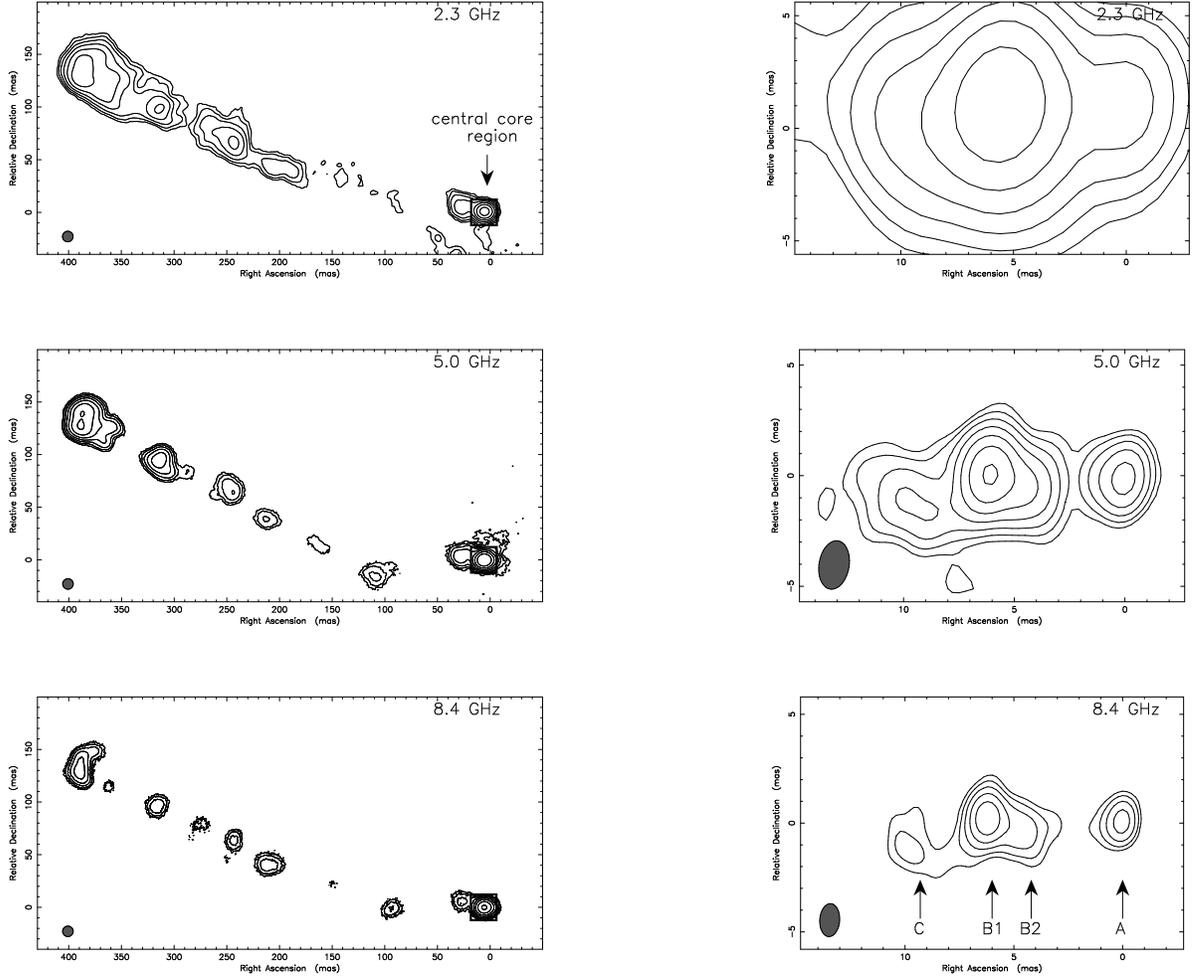}
\caption{Images of 3C 138 at 2.3 GHz (upper panel), 5.0 GHz
(middle panel), and 8.4 GHz (lower panel). On the left are large
scale images which are restored with the same 10 mas circular beam
and, have peak flux densities of 319, 210 and 149 mJy beam$^{-1}$
at 2.3, 5.0 and 8.4 GHz, respectively. On the right are full
resolution VLBA images of the central core region (corresponding
to the blocked area shown in the left) with the elliptical beam
sizes and peak flux densities of 5.1$\times$3.1 mas at --9$^{\rm
o}$ and 202 mJy beam$^{-1}$, 2.2$\times$1.4 mas at --9$^{\rm o}$
and 97 mJy beam$^{-1}$, and 1.5$\times$0.9 mas at --4$^{\rm o}$
and 62 mJy beam$^{-1}$ at 2.3, 5.0 and 8.4 GHz, respectively. The
off source rms noise levels ($\sigma$) are 6.7, 2.7 and 4.1 mJy
beam$^{-1}$ at 2.3, 5.0 and 8.4 GHz, respectively. Contour levels
are drawn at 5$\sigma$$\times$(--1, 1, 2, 4, 8, 16, 32, 64).
\label{fig1}}
\end{figure}

\begin{figure}
\includegraphics[angle=-90,scale=.70]{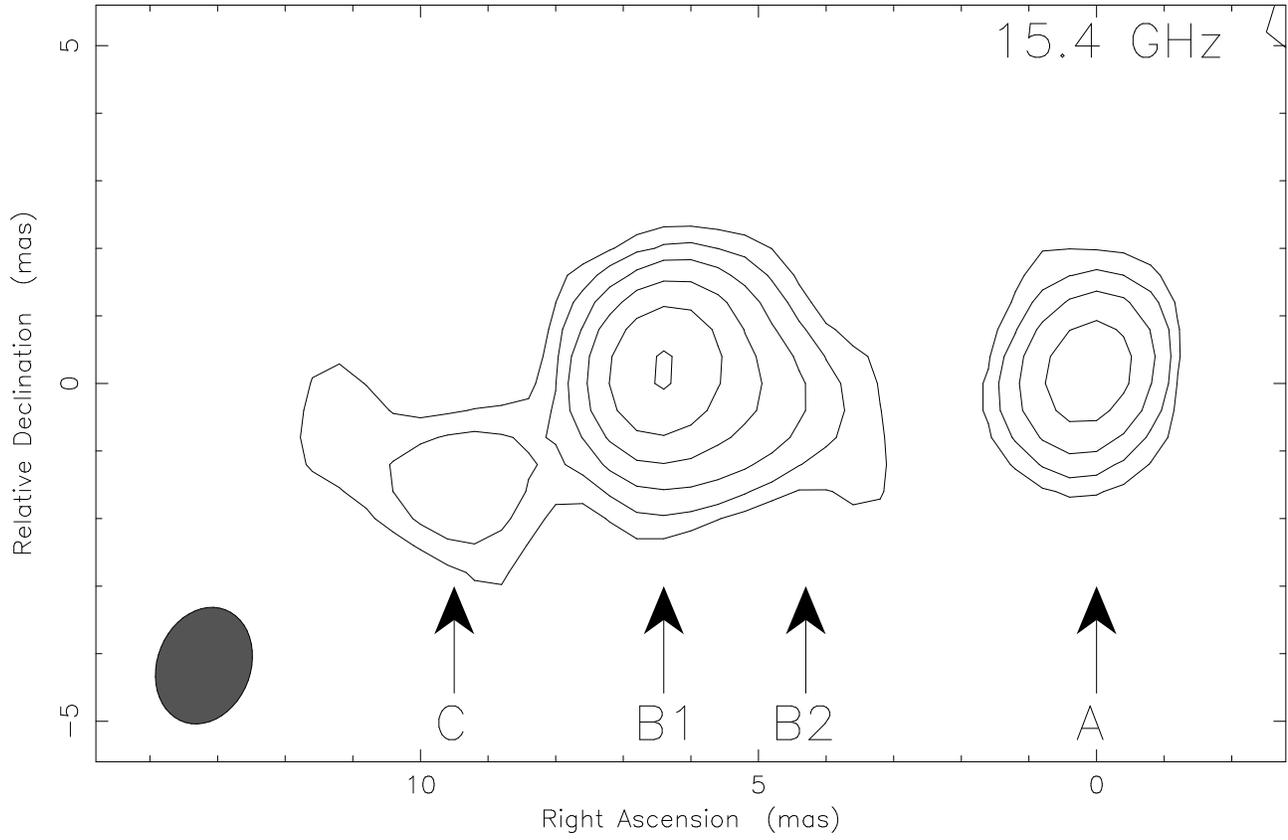}
\caption{High-resolution 15.4 GHz VLBA image of the central core
region of 3C 138. The resolution is 1.8 mas $\times$ 1.4 mas at
--21$^{\rm o}$. The peak flux density is 61 mJy beam$^{-1}$ and
the rms noise is 0.6 mJy beam$^{-1}$. Contour levels are
3$\sigma$$\times$(1, 2, 4, 8, 16, 32). \label{fig2}}
\end{figure}

\clearpage

\begin{figure}
\plotone{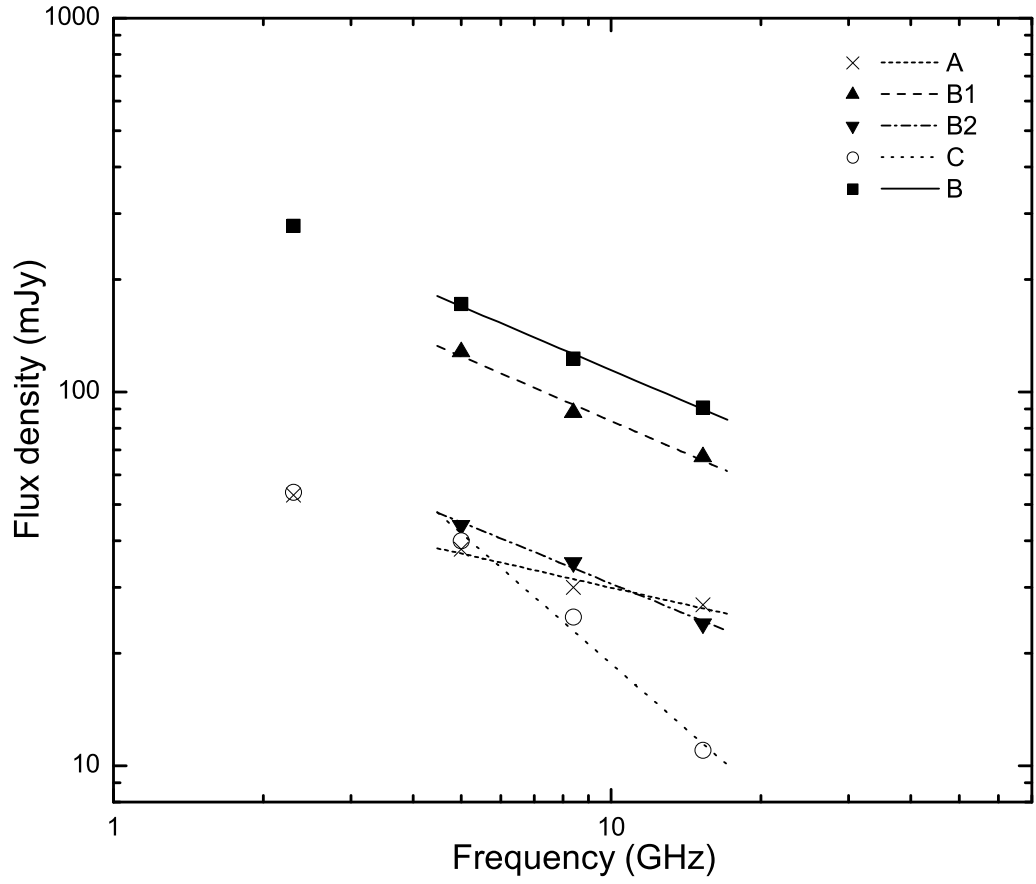} \caption{Spectra of compact components A, B1, B2,
C and B(=B1+B2, see text). Component A has the flattest spectral
index (see Table 2). \label{fig3}}
\end{figure}

\begin{figure}
\plotone{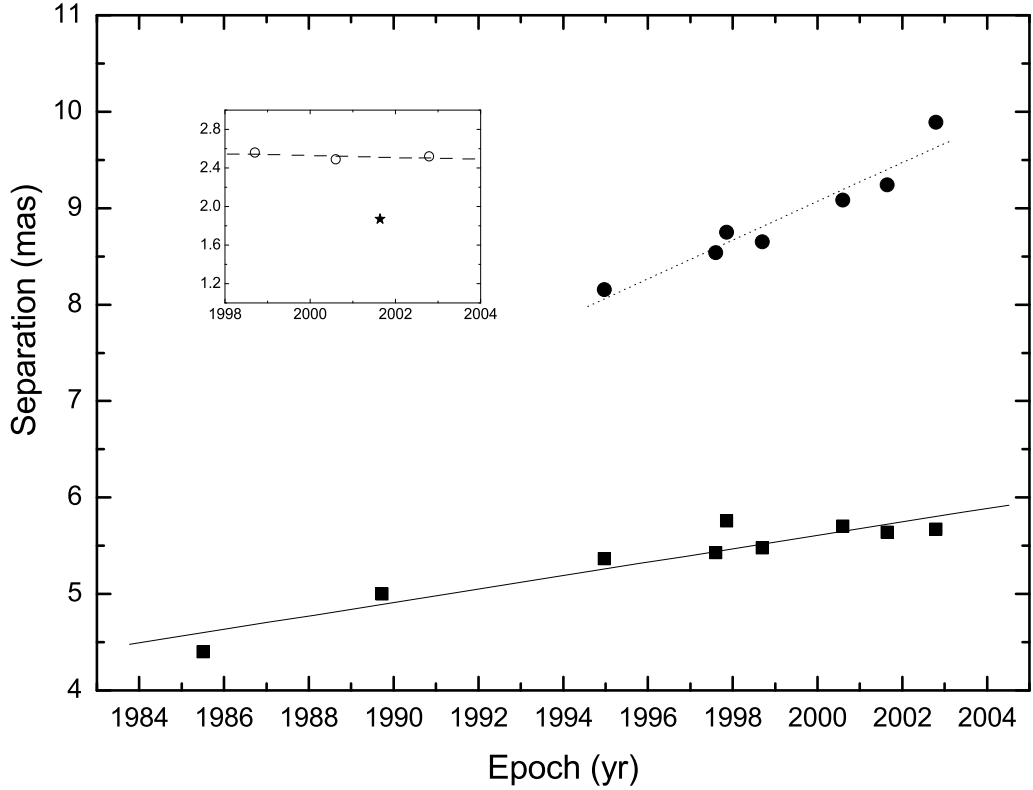} \caption{Separations of component B (filled
squares) and component C (filled circles) relative to component A.
Two lines, solid line and dotted line, represent the best fitted
proper motions of 0.072$\pm$0.010 and 0.20$\pm$0.03 mas yr$^{-1}$
for components B and C, respectively. Inset is a plot of
separation between components B1 and B2 as a function of the
observing epoch. Also shown is a fitting line (dashed line) with a
slope of --0.009$\pm$0.015 mas yr$^{-1}$ (see text). All these
measurements were made from total intensity data at 5.0 GHz.
 \label{fig4}}
\end{figure}

\clearpage

\begin{deluxetable}{lccccl}
\tablecolumns{6} \tablewidth{0pt} \tablecaption{Gaussian model
components in the central region of 3C 138} \tablehead{
\colhead{Component} & \colhead{S (Jy)} & \colhead{r (mas)} &
\colhead{$\theta$ ($^{\rm o}$)} & \colhead{a (mas)} &
\colhead{T$_{\rm B} (\rm K)$}} \startdata \multicolumn{6}{c}{$\nu$\tablenotemark{a} = 2.3 GHz} \\
[3pt] \tableline
\\ [-10pt]
A & 0.053 & 0.00 & ...  & 2.33  & 4.0$\times10^9$ \\
B & 0.278 & 5.19 & 88.5 & 2.35 & 2.0$\times10^{10}$ \\
C & 0.054 & 9.47 & 94.0 & 2.50  & 3.5$\times10^9$ \\
\cutinhead{$\nu$\tablenotemark{a} = 5.0 GHz}
A & 0.038 & 0.00 & ... & 0.94  & 3.7$\times10^9$\\
B2 & 0.044 & 4.30 & 96.0 & 1.36 & 2.0$\times10^9$ \\
B1 & 0.128 & 6.01 & 87.9 & 0.99 & 1.1$\times10^{10}$ \\
C & 0.040 & 9.43 & 96.7 & 2.80 &  4.4$\times10^8$ \\
\cutinhead{$\nu$\tablenotemark{a} = 8.4 GHz}
A & 0.030 & 0.00 & ... & 0.43 & 4.9$\times10^9$ \\
B2 & 0.035 & 4.58 & 94.5 & 1.27 & 6.6$\times10^8$ \\
B1 & 0.088 & 6.20 & 88.4 & 0.70 & 5.4$\times10^9$ \\
C & 0.025 & 9.24 & 97.0 & 2.75 & 1.0$\times10^8$ \\
\cutinhead{$\nu$\tablenotemark{a} = 15.4 GHz}
A & 0.027 & 0.00 & ... & 0.27 & 3.4$\times10^9$ \\
B2 & 0.024 & 4.93 & 94.3 & 1.62 & 8.3$\times10^7$ \\
B1 & 0.067 & 6.36 & 89.9 & 0.44 & 3.2$\times10^8$ \\
C & 0.011 & 9.51 & 99.6 & 1.47 & 4.7$\times10^7$ \\
\enddata

\tablecomments{S: the flux density in Jy; (r,$\theta$): the
distance and position angle of each component with respect to
component A in mas and degrees, respectively; a: the diameter
(FWHM) of circular Gaussian component in mas; T$_{\rm B}$: the
brightness temperature in the source rest frame in K}

\tablenotetext{a}{the frequency corresponding to the following
model components}

%% You can append references to a table using the \tablerefs command.

\end{deluxetable}

\clearpage

\begin{deluxetable}{lccccc}
\tablecolumns{5} \tablewidth{0pt} \tablecaption{Fitted spectral
indices of the central compact components in 3C 138}
\tablehead{\colhead{Component} & \colhead{A} & \colhead{B
(=B1+B2)} & \colhead{B2} & \colhead{B1} & \colhead{C} } \startdata
$\alpha_{\rm 5.0 \phd GHz}^{\rm 15.4 \phd GHz}$ & 0.30$\pm$0.08 & 0.56$\pm$0.04 & 0.54$\pm$0.05 & 0.57$\pm$0.08 & 1.15$\pm$0.13  \\
$\alpha_{\rm 2.3 \phd GHz}^{\rm 15.4 \phd GHz}$ & 0.37$\pm$0.05 & 0.59$\pm$0.02 & ... & ... & ...  \\
\enddata
\end{deluxetable}

%% Tables may also be prepared as separate files. See the accompanying
%% sample file table.tex for an example of an external table file.
%% To include an external file in your main document, use the \input
%% command. Uncomment the line below to include table.tex in this
%% sample file. (Note that you will need to comment out the \documentclass,
%% \begin{document}, and \end{document} commands from table.tex if you want
%% to include it in this document.)

%% \input{table}

%% The following command ends your manuscript. LaTeX will ignore any text
%% that appears after it.

\end{document}